%% file: main.tex
\documentclass[sigconf,nonacm]{acmart}

\input{def}

\setcopyright{none}
\renewcommand\footnotetextcopyrightpermission[1]{}

\title{\system: An OS Runtime for Embodied Agents}

\author{Guojun Chen}
\email{guojun.chen@yale.edu}
\affiliation{%
  \institution{Yale University}
  \country{}
}

\author{Alex Schott}
\email{a.schott@yale.edu}
\affiliation{%
  \institution{Yale University}
  \country{}
}

\author{Lin Zhong}
\email{lin.zhong@yale.edu}
\affiliation{%
  \institution{Yale University}
  \country{}
}

\begin{document}

\begin{abstract}
Large language models (LLMs) can plan behavior for embodied agents from natural language, but treating the LLM as a request/response oracle on the critical path is fundamentally at odds with real-time control and concurrent goals.
We argue for an \emph{operating-system-style runtime} for embodied agents, and instantiate this idea in an early prototype, \system.
\system structures LLM-based planning as asynchronous loops at multiple timescales that overlap with execution, and manages the agent's physical body like an OS manages hardware: the \emph{Skill Kernel} arbitrates typed physical subsystems among concurrent per-task \emph{processes}, a scheduler preempts them and resumes or replaces each by source, and \emph{speculative skill streaming} hides LLM latency behind ongoing motion, while a fast first-action path yields visible feedback within a second. Users program behavior through natural language \emph{prescriptions} that \system dispatches to the LLM-based planners or compiles into low-latency interrupt handlers.
Our prototype of Kalos, a Unitree Go2 quadruped, provides preliminary evidence for the design: in our current task suite, it cuts per-step delay by 50\% over step-by-step planning and time-to-first-action by 73\% over monolithic planning, while admitting concurrent tasks at low scheduling overhead.
\end{abstract}
\maketitle

\input{sections/intro}
\input{sections/related}
\input{sections/overview}

\input{sections/programming}

\input{sections/system}

\input{sections/system-hierarchy}
\input{sections/implementation}
\input{sections/evaluation}
\input{sections/limitation}

\bibliographystyle{IEEEtran}
\bibliography{bib/abr-long,bib/typego}

\end{document}

%% file: def.tex
\usepackage[english]{babel}

\usepackage{amsmath}
\usepackage{graphicx}
\usepackage{array}
\usepackage{colortbl}
\usepackage{xurl}
\usepackage{color}
\usepackage{soul}
\usepackage{siunitx}
\usepackage{listings}
\usepackage{subcaption}
\usepackage{float}
\usepackage[normalem]{ulem}
\usepackage{setspace}
\usepackage{placeins}
\usepackage{afterpage}
\usepackage{hhline}
\usepackage{enumitem}
\usepackage{pifont}
\usepackage{arydshln}
\usepackage{adjustbox}
\usepackage{lipsum}
\usepackage{tabularx}
\usepackage{makecell}
\usepackage{wrapfig}
\usepackage{moreverb}
\usepackage{multirow}
\usepackage{xspace}
\usepackage{epsfig}
\usepackage{endnotes}
\usepackage{url}
\usepackage{fancyhdr}
\usepackage{cascadia-code}
\usepackage{xcolor}
\usepackage{caption}
\usepackage[framemethod=TikZ]{mdframed}
\definecolor{codegray}{RGB}{128, 128, 128}
\definecolor{backcolor}{RGB}{250, 250, 250}
\definecolor{rulecolor}{RGB}{120, 120, 120}
\definecolor{keywordcolor}{RGB}{218, 75, 88}
\definecolor{stringcolor}{RGB}{64, 135, 39}
\definecolor{identifiercolor}{RGB}{50, 50, 50}

\usepackage{tikz}
\usetikzlibrary{automata, positioning}

\usepackage{DejaVuSansMono}
\lstdefinestyle{mystyle} {
    commentstyle=\color{rulecolor},
    keywordstyle=\color{keywordcolor},
    stringstyle=\color{stringcolor},
    basicstyle=\ttfamily\scriptsize,
    identifierstyle=\color{identifiercolor},
    backgroundcolor=\color{backcolor},
    breakatwhitespace=false,
    breakindent=0pt,
    breaklines=true,
    captionpos=b,
    keepspaces=true,
    numbers=left,
    numberstyle=\ttfamily\tiny\color{rulecolor},
    numbersep=3pt,
    showspaces=false,
    showstringspaces=false,
    showtabs=false,
    framexleftmargin=0pt,
    frame=lines,
    rulecolor=\color{rulecolor},
    rulesepcolor=\color{gray},
    xleftmargin=0pt,
    xrightmargin=0pt,
}
\usepackage{minted}
\setminted{
  fontsize=\scriptsize,
  bgcolor=backcolor,
  linenos=true,
  numbersep=3pt,
  frame=lines,
  rulecolor=\color{rulecolor},
  framesep=2pt,
  breaklines=true,
  breakanywhere=false,
  breakindent=0pt,
  xleftmargin=0pt,
  xrightmargin=0pt,
  showspaces=false,
  showtabs=false,
}
\definecolor{blueframecolor}{RGB}{173,212,251} 
\definecolor{textgray}{gray}{0.5}
\definecolor{typeflyblue}{RGB}{95,148,247} 
\definecolor{typeflyred}{RGB}{202,85,92}
\definecolor{typeflygreen}{RGB}{82,133,54}
\definecolor{typeflypurple}{RGB}{189,21,240}

\newmdenv[  
  linecolor=blueframecolor,
  outerlinewidth=0.5pt, 
  roundcorner=3pt, 
  backgroundcolor=blueframecolor!10,
  frametitlerule=true,
  innertopmargin=3pt, 
  innerbottommargin=3pt,
  font=\small
]{customframe}

\newcommand{\lstref}[1]{Listing~\ref{#1}}

\definecolor{lightgray}{gray}{0.9}
\definecolor{lightblue}{rgb}{0.9,0.9,1}
\definecolor{blue_bg}{rgb}{0.85,0.85,1}
\definecolor{lightyellow}{rgb}{1,1,0.8}
\definecolor{lightpurple}{rgb}{1,0.85,1}
\definecolor{red}{rgb}{1,0,0}
\definecolor{darkgreen}{rgb}{0.4,0.7,0.3}

\usepackage{tcolorbox}

\newcommand{\code}[1]{\texttt{\footnotesize #1}}

\definecolor{lightpurple}{rgb}{1,0.85,1}

\newcommand{\remove}[1]{}

\definecolor{cyan10}{HTML}{E5F6FF}
\definecolor{cyan20}{HTML}{BAE6FF}
\definecolor{cyan60}{HTML}{0072c3}
\definecolor{cyan70}{HTML}{00539a}
\definecolor{teal10}{HTML}{D9FBFB}
\definecolor{teal20}{HTML}{9EF0F0}
\definecolor{teal60}{HTML}{007d79}
\definecolor{orange10}{HTML}{FFF2E8}
\definecolor{orange20}{HTML}{FFD9BE}
\definecolor{orange60}{HTML}{ba4e00}
\definecolor{blue10}{HTML}{EDF5FF}
\definecolor{blue20}{HTML}{D0E2FF}
\definecolor{magenta10}{HTML}{FFF0F7}
\definecolor{magenta20}{HTML}{FFD6E8}
\definecolor{magenta30}{HTML}{ffafd2}
\definecolor{magenta50}{HTML}{ee5396}
\definecolor{magenta60}{HTML}{d02670}
\definecolor{magenta70}{HTML}{9f1853}
\definecolor{purple10}{HTML}{F6F2FF}
\definecolor{purple20}{HTML}{E8DAFF}
\definecolor{purple30}{HTML}{d4bbff}
\definecolor{purple70}{HTML}{8a3ffc}
\definecolor{rose10}{HTML}{FCF2ED}
\definecolor{rose20}{HTML}{F9D9D1}
\definecolor{red10}{HTML}{FFF1F1}
\definecolor{red20}{HTML}{FFD7D9}
\definecolor{green10}{HTML}{DEFBE6}
\definecolor{green20}{HTML}{A7F0BA}
\definecolor{yellow10}{HTML}{fcf4d6}
\definecolor{yellow20}{HTML}{fddc69}
\definecolor{gray20}{HTML}{e0e0e0}
\definecolor{gray30}{HTML}{c6c6c6}
\definecolor{gray40}{HTML}{a8a8a8}
\definecolor{gray80}{HTML}{393939}

\definecolor{carbon-gray-10}{cmyk}{0.0, 0.0, 0.0, 0.04, 1.00}
\definecolor{carbon-gray-90}{cmyk}{0.0, 0.0, 0.0, 0.85, 1.00}

\newcommand{\system}[0]{TypeGo\xspace}
\newcommand{\name}[0]{\textsf{Kalos}\xspace}

\newcommand{\skmg}[0]{Skill Kernel\xspace}

%% file: sections/intro.tex
\section{Introduction}
\label{sec:intro}
Robots today execute a rich catalog of low-level skills: quadrupeds walk, trot, climb stairs, and balance, while humanoids add grasping and manipulation on top; many platforms ship turnkey APIs for locomotion, speech, and perception. These skills solve basic motion and action well, but they do not tell a robot what to do when a human asks for something new, the scene changes unexpectedly, or several goals compete for attention. The hard problem has shifted from how a robot body \emph{moves} to how an embodied system \emph{decides}, and hard-coded task logic or per-task trained policies do not scale to the open world.


LLMs offer a path to close the decision-making gap: pretrained on large-scale data, they supply the common sense to interpret a situation and compose the robot's existing skills into a plan~\cite{chen2025typefly,codeaspolicies2022}. But using an LLM to drive an embodied agent raises three system challenges that persist even as models become faster and more capable.

(\textit{i}) First, LLM inference is too slow for real-time reaction. There are systems that place a language or vision--language model in the perception--action loop for high-level reasoning to achieve impressive open-world generality~\cite{ahn2022saycan, huang2023voxposer,
yenamandra2023homerobot, liu2024ok, shah2025bumble}. But they operate far below real-time, with long pauses between consecutive actions, so demonstrations are commonly shown at increased playback speed. Small models respond fast but plan poorly, while state-of-the-art models are capable but substantial, adding seconds of latency. This latency--quality tradeoff is fundamental, and neither extreme alone suffices for real-time control.
(\textit{ii}) Second, real-world embodied agents rarely run one task in isolation. It may be patrolling an area while communicating with a nearby human, and still needs to watch for hazards. Arbitrating the robot's shared actuators among concurrent tasks while honoring their priorities and preempting cleanly are classic OS problems that the LLM literature has largely sidestepped.
(\textit{iii}) Third, one-shot natural-language instructions cover only goals. Embodied agents also need \emph{reactive} rules (e.g., ``back up if anything show abruptly within a near space ($0.2$m)'') that must fire far faster than an LLM call. Making natural language a first-class medium for both goals and fast reaction rules, authored by non-programmers without a code editor, remains an open question.

To tackle these challenges, we hypothesize that \emph{LLM-based control of an embodied agent should run as a continuously executing, asynchronous \emph{runtime} rather than a request/response loop around the LLM.} In effect, it works like a small OS for the agent. We build \textbf{\system}, a prototype that realizes this runtime design. \system runs four concurrent agent loops at different timescales over a \emph{Skill Kernel} that mediates the body's actuators. 
\system is orthogonal to work that accelerates a single LLM call. The design instead aims to \emph{hide} LLM calls from the critical path through speculation and tiered dispatch, so any per-call speedup composes directly with its architecture.

This paper develops three mechanisms that make the runtime hypothesis concrete:
\begin{itemize}[leftmargin=1em, itemsep=2pt, parsep=0pt, topsep=2pt]
  \item \textbf{Multi-cadence asynchronous planning (\S\ref{sec:hierarchy}).} \system runs its planning layers as independent loops at distinct levels of abstraction: a fast reflex layer (S0), an action streamer (S1), slower task-decomposer (S2) and scheduler (S3). Specifically, the S1 features \emph{speculative skill streaming}, which overlaps planning with skill execution and stores the upcoming skill calls in a bounded queue. As a result, the LLM planning latency is hided behind ongoing motion while bounded look-ahead preserves per-step adaptivity with dynamic environment.
  \item \textbf{OS-style runtime with semantic scheduling (\S\ref{sec:system}).} 
  Each task creates a process and assigns it a process control block (PCB). A \emph{Skill Kernel} arbitrates the categorized robot's resources for processes. 
  S3 schedules these processes semantically rather than by fixed numeric priority. When two processes compete for the same resource, S3 decides which wins based on the scene and prompt guidelines. Whether a preempted process then resumes depends on its origin: a user task does not resume by default, whereas an S3-initiated task resumes automatically.

  
  \item \textbf{Natural language as a first-class programming medium (\S\ref{sec:programming}).} Users author \emph{prescriptions}, including natural-language \emph{tasks} that specify goals and \emph{reflexes} that specify reactive rules. Tasks are supplied online to the S1--S3 planners, while reflexes compile into Python condition/action functions that S0 evaluates at high frequency. Together, they let non-programmers author both deliberate and reactive behavior without writing code.
\end{itemize}

We report an early implementation of \system (\S\ref{sec:impl}) and use a prototype deployment on a Unitree Go2 quadruped to validate the design direction in single- and concurrent-task scenarios (\S\ref{sec:eval}). In our current task suite, \system cuts per-step delay by 50\% over step-by-step planning and TTFA by 73\% over monolithic planning, while adding modest coordination overhead: about 200\,ms for S3 scheduling and 50\,ms for S0 reflex handling. 

%% file: sections/related.tex
\section{Related Work}
\label{sec:related}

\textit{Hierarchical robot control.}
\system's planner hierarchy is inspired by the classical three-layer architecture~\cite{gat1998three}: where a deliberator forms a plan, a sequencer turns it into actions, and a controller runs primitives. 
\system modernizes this template in three ways. First, it drives the deliberator (S2) and sequencer (S1) with LLMs rather than symbolic planners. Second, it adds two layers the classical template lacks: a reflex layer (S0) for low-latency reaction and a task scheduler (S3) for multi-task arbitration. Third, all layers run asynchronously, so the sequencer emits actions without blocking on the deliberator.
Inspired by dual-process theory~\cite{kahneman2011thinking}, some recent planners pair a fast ``System 1'' with a slow ``System 2'' and route between them~\cite{saha2025system1x,ziabari2025reasoning,zhu2024language,wang2025hierarchicalreasoningmodel}. Such two-tier designs activate only one tier at a time and switch between modes.
In contrast, \system runs its four loops concurrently across timescales (S0--S3), rather than as alternatives selected per request.

\textit{LLM-based robot planning.}
LLMs are widely used to turn instructions into robot plans~\cite{zeng2023llm,jeong2024survey}. Existing systems are either \emph{monolithic planners} that prompt once for a full plan (Code-as-Policies~\cite{codeaspolicies2022}, TypeFly~\cite{chen2025typefly}) or \emph{step-by-step planners} that query per action (SayCan~\cite{ahn2022saycan}, Inner Monologue~\cite{huang2022inner}, ReAct~\cite{yao2023react}). A monolithic plan delays the first action and cannot recover when the scene changes; a step-by-step planner reacts but stalls the robot between actions while it rebuilds its prompt. \system explores a middle point: asynchronous LLM loops across levels of abstraction, plus a speculative planner (S1) that aims to hide planning latency at runtime. Orthogonally, end-to-end vision--language--action models map perception directly to low-level control~\cite{brohan2023rt2}; such a policy can sit \emph{inside} a \system skill, while \system addresses concurrency, preemption, and latency-hiding above the policy level.

\textit{Multi-agent LLM systems.}
Prior multi-agent work arranges agents in chains or graphs: some refine an instruction top-down~\cite{prakash2023llm,kienle2025lodge}, while Mixture-of-Agents~\cite{wang2024mixture} combines outputs from multiple proposer models through an aggregator. In these designs agents run sequentially, causing execution time for a single request to grow with the number of agents. \system instead runs each planner in its own loop, coordinated through synchronization and scheduling rather than sequential dependency, keeping planning latency low.

\textit{Operating systems for LLM agents.}
A recent line of work borrows OS structure for software agents. AIOS~\cite{mei2024aios} adds an ``LLM kernel'' that schedules and manages agent requests, and MemGPT~\cite{packer2023memgpt} treats the context window like virtual memory, paging information in and out. \system shares their OS framing but targets a fundamentally different workload: a multi-tasking \emph{embodied} agent, whose resources include physical actuators with exclusivity and safety constraints rather than context tokens or API calls. 


%% file: sections/overview.tex
\section{System Overview}
\label{sec:overview}

\begin{figure}[t]
  \centering
    \includegraphics[width=0.7\linewidth]{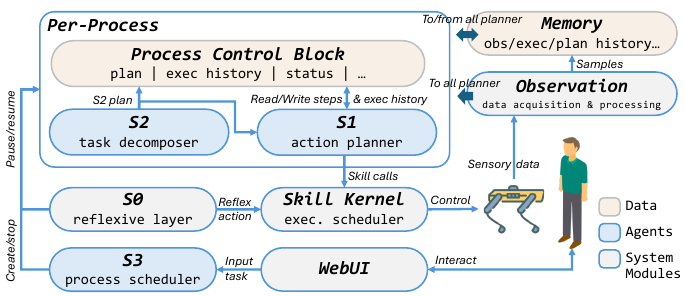}
    \caption{\textmd{\small \system runtime. A hierarchy of agents (blue) includes the global S3 (process scheduler) and S0 (reflexive layer), and the process-scoped S2 (task decomposer) and S1 (action planner). S3 manages the process lifecycle based on user input or its own decisions. S2 and S1 together turn a single task into a continuous stream of skill calls. The \skmg arbitrates and schedules skill calls issued by multiple processes. Observation taps the robot body's sensor stream, and Memory periodically samples the agent's state into a retrieval-accessible store; both are available system-wide.}}
    \label{fig:process-diagram}
\end{figure}

\system realizes the runtime thesis (\S\ref{sec:intro}) as a stack of asynchronous planning loops (S0 through S3) over a \emph{Skill Kernel} that mediates the embodied agent's physical actuators (\autoref{fig:process-diagram}).

The runtime borrows its structure from OS design (\autoref{tab:os-analogy}), which we use as a guiding analogy rather than flavor: most components have an OS counterpart, and we are explicit about where it deliberately departs.
A user-authored \emph{task} is a static, natural-language job specification, analogous to an OS program; the runtime realizes each task as a \emph{process} with its own process control block (PCB). The \emph{Skill Kernel} plays the role of the OS kernel, arbitrating the robot body's typed actuator subsystems (its devices) among concurrent processes; \emph{reflexes} act as interrupt handlers with priority preemption; and the speculative \emph{S1 step queue} resembles a write-behind look-ahead buffer that decouples planning latency from execution.


\begin{table}[t]
\centering
\small
\caption{\textmd{OS analogues used throughout the paper.}}
\label{tab:os-analogy}
\setlength{\tabcolsep}{6pt}
\begin{tabular}{ll}
\toprule
\textbf{\system concept} & \textbf{OS analogue} \\
\midrule
Task (NL job spec, static)      & Program (job specification) \\
Process (handles one task)      & Process \\
PCB (planning state)            & Process control block \\
Skill Kernel                    & OS kernel (resource manager) \\
Subsystem                       & Device with exclusivity policy \\
Skill                           & Driver call \\
Reflex (S0)                     & Interrupt handler \\
S1 step queue                   & Write-behind / look-ahead buffer \\
Source-dependent resumption (S3) & Process scheduler policy \\
Memory (future work)            & File system / virtual memory \\
\bottomrule
\end{tabular}
\end{table}

\paragraph{Where the analogy departs.}
The OS framing is productive precisely because an embodied agent breaks several assumptions a classical OS takes for granted, and these departures are where the research problems lie. First, \emph{resources are physical and semantic at once}: preempting an actuator has real-world, sometimes unsafe, consequences, so clean preemption is a correctness property enforced by bounded-interruption skills. Second, \emph{scheduling requires understanding meaning}: S3 decides which process runs from an LLM's reading of the current observation and behavioral guidelines, a semantic policy a CPU scheduler never needs; additionally, whether a preempted process later resumes depends on the \emph{source} of the preemption rather than a numeric priority (\S\ref{sec:process}). Third, \emph{programs are natural language}: prescriptions are ambiguous and are compiled or interpreted at runtime by LLMs rather than ahead of time (\S\ref{sec:programming}). Fourth, \emph{correctness is probabilistic}: any planning layer can be wrong, so the runtime must contain and recover from bad outputs rather than assume correct programs. We return to these departures throughout.

The three mechanisms behind this runtime map to the sections that follow: \emph{speculative skill streaming} overlaps planning with execution to hide LLM latency (\S\ref{sec:hierarchy}); the Skill Kernel and S3 scheduler together multiplex concurrent tasks, running non-contending processes in parallel and preempting the rest cleanly (\S\ref{sec:system}); and \emph{prescriptions} make natural language a first-class programming medium, routed to the online planners or compiled into S0 interrupt handlers. Users and developers interact with \system through three programming abstractions: \emph{prescription}, \emph{observation}, and \emph{skills}.

%% file: sections/programming.tex
\section{Programming Abstractions}
\label{sec:programming}

\system exposes three programming abstractions: user writes \emph{prescriptions}, and system developers supply \emph{observations} and \emph{skills}.

\input{sections/system-skill-method}

%% file: sections/system-skill-method.tex
A \textbf{prescription} is \system's natural-language programming interface: it lets a user shape how the embodied agent behaves without writing code. The key design point is that \system routes each prescription to the layer whose latency matches its purpose. \system supports two kinds. \textit{Tasks} are high-level objectives (e.g., ``patrol the hallway to find a sports ball'') consumed \emph{online} by the S1--S3 planners; each task spawns its own process, and multiple tasks may coexist. \textit{Reflexes} are NL condition--action pairs that an LLM compiles \emph{offline} into a pair of Python functions over authored skills and observation accessors; S0 evaluates them at high frequency for low-latency reactions (e.g., ``when anything is closer than $0.2$\,m in front, step back $0.5$\,m''). S0 has the highest priority in \system: a firing reflex interrupts any in-progress process holding contended subsystems and resumes it on completion.

\begin{listing}
\centering
\begin{lstlisting}[language=Python,frame=single,basicstyle=\scriptsize\ttfamily]
@robot_skill("name", desc="...", subsystem=SubSystem.MOVEMENT)
def name(arg: type, [pause_evt], [stop_evt]) -> SkillReturn:
    # skill logic; check events for preemption
\end{lstlisting}
\vspace{-2ex}
\caption{\textmd{\system skill interface. A decorator declares the skill's name, description, and subsystem (e.g., \code{MOVEMENT}, \code{SOUND}, \code{DEFAULT}). A skill may optionally accept \code{pause\_evt} and \code{stop\_evt} handles for interruption; skills that omit them are atomic. It returns a \code{SkillReturn} carrying a success flag and message.}}
\label{code:skill_interface}
\end{listing}

An \textbf{observation} is a structured, continuously updated representation of the embodied agent's environment and internal state, grounded in the robot body's sensors and exposed to all planners. It is a collection of \code{fields}, each with a developer-specified semantic role, data source, post-processing, and serialization. For example, an \code{object\_detections} field binds to the front RGB camera, applies YOLO, and serializes detections as JSON for planner prompts.

\textbf{Skills} are callable primitives that implement concrete operations on the robot body; developers author and verify every skill (\lstref{code:skill_interface}). Three principles guide their design. \emph{Synchronous}: a skill blocks until it returns, matching ordinary function-call semantics, though \system invokes it asynchronously through the \skmg (\S\ref{sec:skmg}). \emph{Interruptible or atomic}: interruptible skills weave \code{pause\_evt} and \code{stop\_evt} into their control loop and stop within a bounded time $t_s$, returning a \code{SkillReturn} that distinguishes success, failure, and interruption; a few hardware-constrained skills (e.g., \code{stand}) are atomic. This bounded-interruption guarantee is what lets \system recover from failures and switch tasks with bounded delay. \emph{Authored, not synthesized}: a skill's control flow is fixed by the developer and never synthesized at runtime by \system's agents, though it may call learned models internally, keeping core operations predictable and verifiable.

Most skill interfaces are platform-agnostic, so \system's planners operate largely independently of the underlying hardware. Composite skills may encapsulate an entire optimized control procedure in a single invocation, provided they honor preemption within $t_s$.

%% file: sections/system.tex
\section{Runtime Abstractions}
\label{sec:system}

\system's runtime is structured like a small operating system for an embodied agent. Two abstractions carry the OS analogy: the \emph{Skill Kernel} (\S\ref{sec:skmg}), the resource manager that arbitrates typed actuator subsystems of the robot body among concurrent processes, and the \emph{process} (\S\ref{sec:process}), a schedulable, stateful unit that handles one user task. Together they meet two design objectives:
(1) decouple skill design from scheduling, giving developers an extensible interface for adding new capabilities without touching the core runtime; and
(2) enable safe concurrent and interruptible execution across multiple tasks, so that processes with disjoint exclusive subsystems run in parallel while conflicting ones are serialized or preempted.

\input{sections/system-task}

%% file: sections/system-task.tex
\subsection{\skmg}
\label{sec:skmg}
The \skmg is \system's resource manager. Like an OS kernel mediating device access among processes, it mediates a typed registry of actuator \emph{subsystems} among the processes. It enforces two properties: across processes, two skill calls cannot conflict over a shared physical resource; within a process, a new skill call may override an in-flight one. The kernel arbitrates by acquiring subsystems on behalf of \emph{skills}, the analogue of driver calls.

\textit{Subsystems.}
A \emph{subsystem} abstracts a physical resource of the robot body that a skill consumes during execution. Each is configured along two axes: \emph{exclusive\_per\_process} (only one process may hold it at a time) and \emph{parallel} (multiple invocations may run concurrently). Developers can add a new subsystem (e.g., a manipulator arm or a display) without modifying the scheduler or existing skills. \system ships three: \code{MOVEMENT} (exclusive, serial) for locomotion and gestures, which rejects a second requester to prevent conflicting motion; \code{SOUND} (shared, serial) for audio output, which serializes invocations in a queue; and \code{DEFAULT} (shared, parallel) for skills needing no exclusive hardware.

\textit{Execution interface.}
When a skill such as \code{move\_forward} is invoked, its declared subsystem (\code{MOVEMENT}) is marked held by the calling process, blocking other processes that need the same exclusive subsystem. Processes use the non-blocking \code{execute} interface, which returns immediately with a unique id and a flag indicating whether the skill was scheduled or rejected due to a conflict; the caller queries progress via the id.

\subsection{Process Abstraction}
\label{sec:process}
A \emph{process} is \system's runtime unit of multitasking, modeled on an OS process. The runtime realizes each task by spawning a process with a private process control block (PCB) which holds the process \emph{ID}; the \emph{task} text (user- or S3-generated); the S2/S1 \emph{plan} (subtask list and index, global events, S1 step queue); the skill-execution \emph{history}; the lifecycle \emph{status} (\textsc{In-Progress}, \textsc{Paused}, \textsc{Success}, \textsc{Failed}); and the robot skill set.

\textit{Life cycle.}
For every incoming task, S3 instantiates a PCB and a dedicated S1/S2 pair. While \textsc{In-Progress}, S2 decomposes the task and S1 streams skill calls; each call attempts to acquire its required subsystems, and any failure moves the process to \textsc{Paused}, where both loops block until S3 reschedules it. The process ends in a terminal state when it completes or is stopped by S3 or the user.

\textit{Process types.}
\system distinguishes processes by \emph{source} rather than numeric priority: \textbf{user} processes carry explicit instructions; \textbf{reactive} processes are spawned by S3 when standing behavioral guidelines fire; and a single \textbf{idle} process runs default behavior whenever nothing else is active.

\textit{Scheduling: concurrency.}
Two processes run simultaneously only when the exclusive subsystems they hold are disjoint; otherwise, one is \textsc{Paused} until the other releases its subsystems. A process retains its subsystems until it pauses or terminates; it keeps \emph{MOVEMENT} reserved even after \code{move\_forward} finishes, because a following \code{take\_pic} may depend on the physical body staying still.

\paragraph{Scheduling: source-dependent resumption.}
\system encodes two post-preemption behaviors that a single numeric priority would conflate. \emph{Interrupt-and-return} fires when a transient event (e.g., a person walks into view) triggers a brief reactive detour: the new process takes over, completes, and the displaced process auto-resumes. \emph{Replace-without-return} fires when a new user instruction supersedes the old goal for good: the previous process is stopped and never auto-resumed. Flat priority schemes tell the runtime which process \emph{wins} but not whether to stash the loser for later or discard it. S3 derives the distinction from the preemption's source: a \textbf{user} process replaces-without-return (S3 stops every running user and reactive process, none auto-resumed unless the user's task explicitly asks for it); a \textbf{reactive} process always interrupts-and-returns (it preempts only user or idle processes, and S3 auto-resumes the predecessor when it terminates); the \textbf{idle} process never preempts.

%% file: sections/system-hierarchy.tex
\section{Hierarchical Collaborative Agents}
\label{sec:hierarchy}
\system pairs rapid response with deliberation through four asynchronous agents, from S3 (slowest, most abstract) down to S0 (fastest, most concrete). The names generalize the System 1 / System 2 dichotomy~\cite{kahneman2011thinking} to four cadences, but, unlike that two-tier view, our tiers run \emph{concurrently} rather than as alternatives selected per request. S3 and S0 are global across processes; S2 and S1 are instantiated per process. Each runs its own independent loop.

\paragraph{System 3: process scheduler}
S3 manages processes from two inputs: explicit user tasks and behavioral guidelines embedded in its prompt. The second lets \system act autonomously. For example, under the guideline ``you may pause a non-emergency task to engage a nearby person,'' S3 pauses ``find a sports ball'' and starts ``play with the person'' when it detects someone. User-initiated processes outrank S3-created ones; S3's concurrency and preemption policy is detailed in \S\ref{sec:process}.

\paragraph{System 2: task decomposer}
S2 decomposes a task into (1) an ordered list of \emph{subtasks}, each a natural-language description of what to achieve plus a success criterion, which S1 consumes one at a time; and (2) a set of \emph{global events}, condition/response pairs that S1 monitors as reactive interrupts during execution (Listing~\ref{lst:s2-plan}). After the initial decomposition, S2 periodically (every 2\,s in our implementation) re-inspects the action history and observation and decides whether to keep or update the decomposition.

\begin{listing}
\centering
\begin{lstlisting}[language=Python,frame=single,basicstyle=\tiny\ttfamily,breaklines=true]
# S2 output for "Find an apple and report to the user. Take a picture whenever you see a ball".
{"operation": "update",
  "subtasks": [
    {"desc": "Find the apple", "criteria": "Found, or nowhere left to search"},
    {"desc": "Report to the user", "criteria": "See and speak to user"}],
  "global_evt": [{"cond": "see a ball", "resp": "take a picture"}]}
\end{lstlisting}
\vspace{-2ex}
\caption{\textmd{\small S2 output for a sample task, decomposed into two subtasks and one global event.}}
\label{lst:s2-plan}
\end{listing}

\paragraph{System 1: action planner}
\label{sec:s1}
S1 turns the current subtask into a stream of skill calls (Listing~\ref{lst:s1-plan}) via two concurrent threads: a \emph{streamer} that generates steps and an \emph{executor} that runs them, both able to call an LLM. Rather than emit all steps at once, the streamer generates a few per LLM call into a bounded queue (size three in our case), while the executor dequeues the oldest; each dequeue triggers one enqueue, so generation overlaps execution. The stream-interpretation idea itself follows TypeFly~\cite{chen2025typefly}; \system's contribution is to embed it in a multi-cadence runtime, where S2 continuously revises the subtask that S1 streams against and S3 can preempt the whole loop, rather than running it as a single open-loop interpreter. Each \emph{step} holds one or more mutually exclusive \emph{branches} pairing a condition (a predicate over observation or a short NL expression) with a skill call; the executor invokes the matching branch. The executor also watches S2's global events: when one fires, it interrupts the in-flight skill, records the event on the PCB, and the streamer handles it on its next cycle.

\begin{listing}
\centering
\begin{lstlisting}[language=Java,frame=single,basicstyle=\tiny\ttfamily,breaklines=true]
// In-progress skill: stand_up(), Subtask: Find the apple.
{"steps": [
    {"branches": [{"action": "search('apple')", "cond": "always"}]},
    {"branches": [
      {"action": "approach('apple')", "cond": "visible('apple')"},
      {"action": "goto_waypoint(3)", "cond": "always"}
    ]}
  ],
  "subtask_complete": false}
\end{lstlisting}
\vspace{-2ex}
\caption{\textmd{\small S1 streamer output for a sample subtask. While \code{stand} runs, the streamer speculatively generates two further steps: search for the apple, and then approach it if visible or move on otherwise.}}
\label{lst:s1-plan}
\end{listing}


\paragraph{System 0: reflexive layer}
S0 evaluates a list of condition functions and invokes the matching action function; both are plain Python that an LLM compiles from reflexes offline, so no LLM sits on the reaction path. S0 outranks every task-driven process: a firing reflex pauses every process holding a contended subsystem and resumes each on completion. Reflexes themselves preempt one another by hard priority.

%% file: sections/implementation.tex
\section{Implementation}
\label{sec:impl}
Our current prototype is implemented in Python and packaged as a ROS2 node, with three parts: the \system runtime (planning and execution), a lightweight platform-specific SDK that exposes skills and sensor data over ROS2, and a pool of AI services (LLMs and perception models). S0 runs at 100\,Hz, S2 and S3 at 0.5\,Hz, and S1 is event-driven.


\textit{Platform SDK and case study (\name).}
The ROS2 SDK is \system's data and control provider for the robot body: it publishes sensor data (RGB images, LiDAR scans, state estimates) and forwards control commands over TCP, and integrates SLAM and navigation with an indexed waypoint service. Our initial deployment, \name, runs \system on a Unitree Go2 and registers skills in four categories: basic locomotion (\code{move\_forward}, \code{rotate}, \code{nav}), etc; postural and expressive actions (\code{stand}, \code{sit}, \code{nod}); high-level composites (\code{search}, \code{approach}, \code{follow}, \code{goto\_waypoint}, \code{patrol}); and system utilities (\code{take\_pic}, \code{speak}, \code{query\_memory}). \system is not tied to the Go2: it runs on platforms as modest as the Petoi quadruped~\cite{opencat_quadruped_robot} or a wheeled robot, requiring only basic locomotion and continuous perception.

\textit{AI services and deployment.}
An edge server (with an RTX 4090) runs an HTTP gateway that forwards LLM calls to the cloud and dispatches perception to a pool of stateless gRPC workers: object detection (YOLO~\cite{redmon2016yolo}), open-vocabulary detection (OmDet~\cite{zhao2024omdet}), image--text embedding (CLIP~\cite{ilharco_gabriel_2021_5143773}), and TTS. Our deployment uses two off-robot machines over Wi-Fi: a laptop for the runtime and a GPU edge server, but the stack can be consolidated onto a single on-robot edge device (e.g., an NVIDIA Jetson) for self-contained mobile operation.

%% file: sections/evaluation.tex
\section{Preliminary Evaluation}
\label{sec:eval}


We use a small prototype task suite to validate three claims from \S\ref{sec:intro}: \textbf{effectiveness} (can the prototype complete composite, long-horizon, perception-grounded tasks end-to-end, and degrade gracefully when it cannot?); \textbf{responsiveness} (does layered planning reduce per-step latency relative to a re-plan-every-step controller while retaining more adaptivity than a monolithic planner?); and \textbf{concurrency} (can the runtime admit concurrent tasks or reflexes, how fast is preemption, and at what cost?). The goal is not to exhaustively benchmark robot planning, but to test whether the proposed OS-style abstractions behave as intended on a real embodiment: effectiveness probes the end-to-end runtime, responsiveness probes multi-cadence planning with speculative streaming, and concurrency probes the Skill Kernel together with scheduling.

\paragraph{Baselines and setup.}
Both baselines share \system's skill library and use minimal-reasoning prompts. \system and both baselines drive every planning layer with GPT-5.4; the sole exception is \system's S1P fast-response path, which uses Groq-hosted GPT-OSS-120B (\S\ref{sec:s1}). \emph{ReAct}~\cite{yao2023react} emits one reasoning step plus one skill call per step from the instruction, current observation, and prior trace; it has no decomposition/execution split, speculation, reflex, or scheduler, so it cannot plan its next action until the current skill returns. \emph{Plan-and-execute (PAE)} prompts once for a complete ordered plan, executed open-loop. We run three task categories (\autoref{tab:task_list}), with 10 trials per task.

\begin{table*}[t]
    \centering
    \caption{\textmd{\small Evaluation tasks for \system on Kalos.}}
    \vspace{-1em}
    \footnotesize
    \begin{tabularx}{\textwidth}{p{0.3cm} >{\raggedright\arraybackslash}p{1.6cm} X}
    \toprule
    \textbf{ID} & \textbf{Category} & \textbf{Task Description} \\
    \midrule
    T1 & \multirow{3}{=}{Explicit Commands} & Turn around and face the person. \\
    \cline{1-1} \cline{3-3}
    T2 & & Step back, turn left twice, and nod three times. \\
    \cline{1-1} \cline{3-3}
    T3 & & Go to the hallway (path blocked; the system should report failure). \\
    \midrule
    T4 & \multirow{3}{=}{Open-ended Planning} & Look for a chair in the hallway and take a picture of it. \\
    \cline{1-1} \cline{3-3}
    T5 & & Look around for a sports ball. If you find one, walk up to it and bark twice. If you've checked the whole area and there's none, return to the start and sit down. \\
    \midrule
    T6 & \multirow{3}{=}{Schedule / Concurrency} & While running T5, the user issues ``stop, check what's on your left, then resume'' --- the embodied agent must abort safely and then continue. \\
    \cline{1-1} \cline{3-3}
    T7 & & While running T5, ask the embodied agent how many places it has explored. \\
    \cline{1-1} \cline{3-3}
    T8 & & While running T5, a person suddenly appears in front of the robot and trips S0's obstacle-avoidance reflex once. \\
    \bottomrule
    \end{tabularx}
    \label{tab:task_list}
\end{table*}

\subsection{Effectiveness and Responsiveness}
\label{sec:eval-micro}

\begin{table*}[t]
\centering
\caption{\textmd{Success count and latency breakdown across planners (T1--T5); total completion time and token usage are normalized before averaging. The results suggest that the runtime can reduce reaction and per-step delay relative to ReAct, at the cost of higher token usage.}}
\vspace{-1em}
\label{tab:micro_benchmark}
\footnotesize
\setlength{\tabcolsep}{7pt}
\begin{tabular}{lcccccccccc}
\toprule
& \multicolumn{5}{c}{Success (/10) $\uparrow$}
& & \multicolumn{3}{c}{Avg.\ Time (mean/std) $\downarrow$} & Tokens $\downarrow$ \\
\cmidrule(lr){2-6} \cmidrule(lr){8-10}
Planner & T1 & T2 & T3 & T4 & T5 & & TTFA (s) & Total (Norm.) & Step (s) & Norm. \\
\midrule
\system & 10 & 10 & \textbf{8} & 6 & \textbf{10} & & \textbf{0.87/0.05} & \textbf{1.00/0.12} & 0.62/0.54 & 1.00/0.33 \\
ReAct   & 10 & 10 & 7 & 6 & 9 & & 1.46/0.23 & 1.33/0.09 & 1.25/0.27 & 0.32/0.08 \\
PAE     & 10 & 10 & 0 & \textbf{9} & \textbf{10} & & 3.27/1.73 & 1.14/0.26 & \textbf{0.01/0.01} & \textbf{0.05/0.02} \\
\bottomrule
\end{tabular}
\vspace{-0.8em}
\end{table*}
\autoref{tab:micro_benchmark} reports success and three latency metrics: \emph{TTFA} (input to first action), \emph{total} (normalized end-to-end time), and \emph{per-step} (wait between skills), plus normalized token usage. In this prototype task suite, \system attains the lowest TTFA (40\% below ReAct, 73\% below PAE) and the lowest completion time, and cuts per-step delay by 50\% relative to ReAct. Two design choices drive this. First, S1P emits one immediate skill call at process creation, overlapping the first action with the higher-level planners' deliberation and pulling reaction latency under one second; this action matches the eventual plan in roughly 60\% of cases but consistently produces visible feedback within the first second. Second, the speculative step queue keeps upcoming skill calls ready, removing the per-step planning stall that dominates ReAct's latency. The cost is token usage: \system's multiple asynchronous loops consume substantially more tokens than ReAct (one loop) or PAE (one initial call), an explicit responsiveness--cost tradeoff we leave to future runtime-policy optimization.

\paragraph{Failure cases.}
T3 has a blocked path: \system and ReAct both recognize the failure and report it after a few attempts, while PAE has no closed-loop feedback and stalls indefinitely (0/10). This supports the need for reactive re-planning, but also shows why the prototype should be tested on richer failure modes. T4 is the only task where PAE wins: its global plan commits to going to the hallway before searching, whereas \system and ReAct greedily inspect chairs already in view. It is a known bias of step-by-step planning toward locally promising actions~\cite{wang2026reasoningfailsplanplanningcentric, xu2026faraheadllmsplan} and an open design point for the S1/S2 split.

\subsection{Concurrency and Scheduling}
\begin{table}[t]
\centering
\caption{\textmd{Scheduling overhead (base task T5). \emph{Add-on} is the concurrent task's standalone duration; \emph{Total} is both tasks together under \system's scheduling. \emph{Overhead} is the runtime beyond running the activities independently: for the serialized path (T6) it splits into S1 step-queue replan plus true scheduling cost ($1.20+0.22$); for the parallel path (T7) the add-on overlaps the base, so overhead is compared with the base alone; T8 is the one-shot S0 reflex detour.}}
\label{tab:task_overhead}
\vspace{-1em}
\footnotesize
\begin{tabular}{lcccc}
\toprule
Task & Base (s) & Add-on (s) & Total (s) & Overhead (s) \\
\midrule
T6 & \multirow{3}{*}{34.76/2.16} & 6.17/1.58  & 42.35/2.37 & 1.20 + 0.22 \\
T7 &                            & 4.37/1.21  & 34.91/1.94 & 0.15 \\
T8 &                            & 0.50/0.01 & 35.31/2.13 & 0.05 \\
\bottomrule
\end{tabular}
\vspace{-1.0em}
\end{table}
Each concurrency task injects a second activity during T5, exercising a different scheduling path and, with it, a different facet of the abstractions. T6 and T8 both follow the \emph{interrupt-and-return} path (one triggered by a user instruction, one by an S0 reflex) and qualitatively confirm that the displaced process is paused and correctly resumed afterward, while T7 confirms that processes holding disjoint subsystems run in parallel. Since neither baseline can admit a second activity mid-execution, we report only \system's scheduling overhead---the runtime beyond running the activities independently: their summed standalone durations when they contend and serialize (T6), or the base alone when the add-on runs in parallel (T7). Overhead tracks subsystem contention (\autoref{tab:task_overhead}). In T6, the injected instruction contends with T5 for the \code{MOVEMENT} subsystem, so S3 serializes the two, pausing and resuming T5; of the 1.42\,s total, roughly 1.2\,s is S1 regenerating its step queue on resume, leaving only a small remainder as true scheduling overhead. T7's memory query uses no movement subsystem and runs in parallel (0.15\,s), and T8's S0 reflex preempts T5 for a single obstacle-avoidance action before immediately returning (0.05\,s). These early results suggest that runtime overhead is tied to subsystem contention, not merely to the presence of a second activity.


%% file: sections/limitation.tex
\section{Future Work: Toward a Full Agent OS}

\system is an early step toward an operating system for embodied agents, not a finished OS. Our prototype focuses on three primitives that were immediately necessary for validating the idea on a physical embodiment: concurrent processes, resource arbitration over shared actuators, and multi-cadence planning. A complete agent OS needs additional primitives that make long-lived, adaptive agents manageable across hours, days, and deployments.

\paragraph{Memory as an OS primitive.}
Memory remains the most important area for future work. Today, \system includes only a simple memory service: the runtime samples observations and execution history, stores them in a retrieval-accessible database, and exposes retrieval through skills such as \code{query\_memory}. This works, but it treats memory as an application service rather than an OS abstraction. In an agent OS, memory should be a first-class managed resource, analogous to a file system or virtual memory: processes should name, share, isolate, evict, checkpoint, and recover state through a common interface.

We envision a memory hierarchy mirroring the planners' cadence, so each layer reads state at the abstraction and frequency it can afford: \textbf{M0}, raw structured observations (detections, pose, timestamps) consumed by the fast S0/S1 loops; \textbf{M1}, natural-language observations distilled from M0 (``at noon, a ball was seen near the kitchen''); \textbf{M2}, episode summaries that compress M1 and human interactions for S2; and \textbf{M3+}, long-term knowledge such as habits and regularities accumulated across tasks and deployments for S3's strategic decisions. Lower tiers update at high frequency near perception; higher tiers are refined by \emph{active consolidation}, asynchronous background processes that group, summarize, and promote entries without stalling real-time control. Crucially, rather than exposing memory only through an explicit query API, the runtime itself would choose which tier to inject into each planner's context, or even into its KV cache, so memory continuously shapes planning the way virtual memory backs computation.

This raises OS questions that are distinct from ordinary vector-store design. What is the unit of memory: a raw sensor frame, an event, an executed skill, a failed plan, a user preference, or a distilled episode? Which memories are process-private, which are shared across processes, and which survive process termination? How should the runtime arbitrate memory bandwidth and storage budget among competing agents? What consistency guarantees should hold when S1, S2, S3, and S0 read or write different views of the same world state? When should observations be evicted from memory, given that semantically obsolete data may remain safety-relevant? These questions are central to the OS vision because planning, scheduling, and resource control all depend on what the agent can remember.

\paragraph{Richer resource and isolation models.}
We evaluated \system on a quadruped robot, whose main exclusive resource is locomotion. Other embodiments, such as humanoids and mobile manipulators, will require richer subsystem declarations: arms, hands, gaze, torso posture, displays, tools, private data streams, and external APIs. The Skill Kernel mechanism already supports new subsystems, but future systems should expose stronger isolation policies: which processes may see which observations, which tools may externalize data, and which skills may run concurrently under safety constraints. This moves the Skill Kernel closer to an OS protection boundary, not just a scheduler for actuators.

\paragraph{Cost-aware runtime policy.}
The current prototype deliberately spends tokens to buy responsiveness. A production agent OS should make this tradeoff explicit and dynamic: S2 monitoring could become event-triggered, S1 could skip re-generation when the queue and observation are unchanged, and the use of smaller distilled models. More generally, the scheduler should consider not only actuator contention but also LLM budget, model latency, network availability, and risk.
